\documentclass[fleqn,10pt]{wlscirep}
\usepackage[utf8]{inputenc}
\usepackage[T1]{fontenc}
\usepackage{placeins}

\title{A Memristor-Based Bayesian Machine}

\author[1,+]{Kamel-Eddine~Harabi}
\author[1,2,+]{Tifenn~Hirtzlin}
\author[1,+]{Cl\'ement~Turck}
\author[2,]{Elisa~Vianello} 
\author[3]{Rapha\"el~Laurent}
\author[3,4]{Jacques~Droulez}
\author[4]{Pierre~Bessi\`ere}
\author[5]{Jean-Michel~Portal}
\author[5]{Marc~Bocquet}
\author[1,*]{Damien~Querlioz}
\affil[1]{Universit\'e Paris-Saclay, CNRS, Centre de Nanosciences et de Nanotechnologies, 91120 Palaiseau, France}
\affil[2]{CEA, LETI, Universit\'e Grenoble-Alpes, 38400 Grenoble, France}
\affil[3]{HawAI.tech, 38400 Grenoble, France}
\affil[4]{Institut des Syst\`emes Intelligents et de Robotique, Sorbonne Universit\'e, CNRS, 75005 Paris, France}
\affil[5]{Institut Mat\'eriaux Micro\'electronique Nanosciences de Provence, Univ. Aix-Marseille et Toulon, CNRS, 13007 Marseille, France}
\affil[*]{damien.querlioz@c2n.upsaclay.fr}
\affil[+]{these authors contributed equally to this work}

\begin{abstract}
In recent years, a considerable research effort has shown the energy benefits of implementing neural networks with memristors or other emerging memory technologies. However, for extreme-edge applications with high uncertainty, access to reduced amounts of data, and where explainable decisions are required, neural networks may not provide an acceptable form of intelligence. Bayesian reasoning can solve these concerns, but it is computationally expensive and, unlike neural networks, does not translate naturally to memristor-based architectures. In this work, we introduce, demonstrate experimentally on a fully fabricated hybrid CMOS-memristor system, and analyze a Bayesian machine designed for highly-energy efficient Bayesian reasoning. The architecture of the machine is obtained by writing Bayes’ law in a way making its implementation natural by the principles of distributed memory and stochastic computing, allowing the circuit to function using solely local memory and minimal data movement. Measurements on a fabricated small-scale Bayesian machine featuring 2,048 memristors and 30,080 transistors show the viability of this approach and the possibility of overcoming the challenges associated with its design: the inherent imperfections of memristors, as well as the need to distribute very locally higher-than-nominal supply voltages. The design of a scaled-up version of the machine shows its outstanding energy efficiency on a real-life gesture recognition task: a gesture can be recognized using 5,000 times less energy than using a microcontroller unit. The Bayesian machine also features several desirable features, e.g., instant on/off operation, compatibility with low supply voltages, and resilience to single-event upsets. These results open the road for Bayesian reasoning as an attractive way for energy-efficient, robust, and explainable intelligence at the edge.
\end{abstract}
\begin{document}
\maketitle
\thispagestyle{empty}


\section*{Introduction}

Bringing intelligence at the edge is an essential goal of electronics research, to empower embedded systems able to  act for people’s health, or monitor the safety of our buildings, our industrial installations, and the environment.
Unfortunately, the artificial intelligence (AI) algorithms able to provide such services consume a considerable amount of energy when  operated on conventional hardware \cite{editorial_big_2018}. 
Therefore, the lack of locally available energy forces most edge systems to upload their sensed data for processing in the cloud, raising privacy and security concerns.  
In-memory or near-memory computing approaches are a major lead to reducing the energy consumption of AI \cite{zhang2020neuro,sebastian2020memory,markovic2020physics}, as these approaches reduce  data movement considerably, which is the dominant source of energy consumption in AI algorithms \cite{pedram2017dark}.
In the literature, this type of approach has been mostly applied to artificial neural networks.
These algorithms feature neurons connected by synapses and are thus highly topological. Therefore, it is natural to implement  them with near-memory computing: memory devices implement synapses, and neuronal computation is implemented as close as possible to the memories \cite{ambrogio2018equivalent,yu2018neuro,prezioso2015training,ielmini2018memory,wang2018fully,xue2021cmos}.
This approach takes particular appeal when using emerging memory technologies (memristors, resistive memory, phase change memory, magnetoresistive...), which can allow integrating massive amounts of non-volatile memory at the core of logic, in the back end of line of CMOS.

Regrettably, for applications such as intelligent medical sensors, neural networks have some important limitations. First, they need to be trained by massive amounts of data, which are often not available \cite{chen2019deep,ghassemi2020review}. Second, their results are non-explainable, which is not acceptable for some critical applications for ethics and regulatory reasons \cite{rai2020explainable}.
By contrast, Bayesian reasoning, or Bayesian inference, is an artificial intelligence approach that could be more adapted to these situations \cite{ghahramani2015probabilistic,letham2015interpretable}.
 Bayesian reasoning is a probabilistic framework that permits decision-making in situations with incomplete information, maximally incorporating all available evidence, assumptions, and prior knowledge \cite{jaynes2003probability,bessiere2013bayesian}. 
 Within this approach, reasoning is fully explainable and excels at ``small data'' situations, as it is able to incorporate prior expert knowledge \cite{van2014gentle}. It can also estimate the certainty of its prediction  \cite{letham2015interpretable}, which is a challenge for neural networks.
 Bayesian models are not directly brain-inspired but have been connected to biological intelligence  \cite{laurens2007bayesian,lee2003hierarchical,maass2014noise,knill2004bayesian,deneve2008bayesian}.

However, although Bayesian reasoning requires considerable memory access, implementing Bayesian reasoning  near-memory is much more challenging than for neural networks. In a Bayesian approach, networks feature a topological nature, but in a way that is much more subtle than neural networks. Bayesian reasoning is usually implemented on conventional computers \cite{smith2020massively}, microcontroller units \cite{leech2017real,lei2020research}, or graphics processing units \cite{ferreira2011bayesian}.
Several works have also implemented it on
large field-programmable gate arrays \cite{zermani2015fpga,cai2018vibnn,liu2016unbiased,frisch2017bayesian}, and CMOS-based ASICs \cite{ko20203mm}. 
However, the energy efficiency of such approaches is always limited by the cost of memory access to the external dynamic random-access memory.

In this work, we show experimentally on a model system incorporating 2,048 memristors and 30,080 transistors that it is possible to implement Bayesian models that collocate  memory based on resistive technology and computation, using a fully fabricated hybrid CMOS/memristor Bayesian machine.
Our architecture uses fully distributed memory (only local memristor arrays are used as memory). 
Due to the locality of computations, and the reliance on the paradigm of stochastic computing, minimum data movement is performed between the different parts of the systems. 
We show energy improvement of several thousand times with regards to a standard implementation of Bayesian inference on a microcontroller unit.
Additionally, due to the use of non-volatile memory, the system has an instant on/off feature: it can perform Bayesian inference almost immediately after being turned on, allowing to cut the power supply entirely as soon as the system is not used. Finally, inference can be performed at  low voltage, and the system is inherently resilient to soft errors (i.e., single-event upsets), making it suitable for systems relying on energy harvesting with unreliable power supply, and working in extreme environments.

Other approaches have connected nanotechnology and  Bayesian inference. 
Several works exploit the stochasticity of nanodevices \cite{faria2018implementing,friedman2016bayesian,vodenicarevic2017low} to allow Bayesian inference. These works do not address how model parameters are stored, while memory access is the dominant source of energy consumption in AI. The results of ~\cite{dalgaty2021situ} also show that the imperfections of  memory devices can be used as a driver for Bayesian learning.  This approach allows the training of simple models and is not adapted to implement possibly more sophisticated models in purely inferential systems.

The paper first introduces the foundational principles of the Bayesian inference machine, then presents our fabricated system and its characterization. Finally, a design study of a scaled-up system on a practical motion recognition task highlights the energy benefits of the Bayesian machine in a real-life setting.


\section*{Results}

\subsection*{Memristor-based stochastic Bayesian machine} 

\begin{figure}[h!]
\centering
\includegraphics[width=0.7\linewidth]{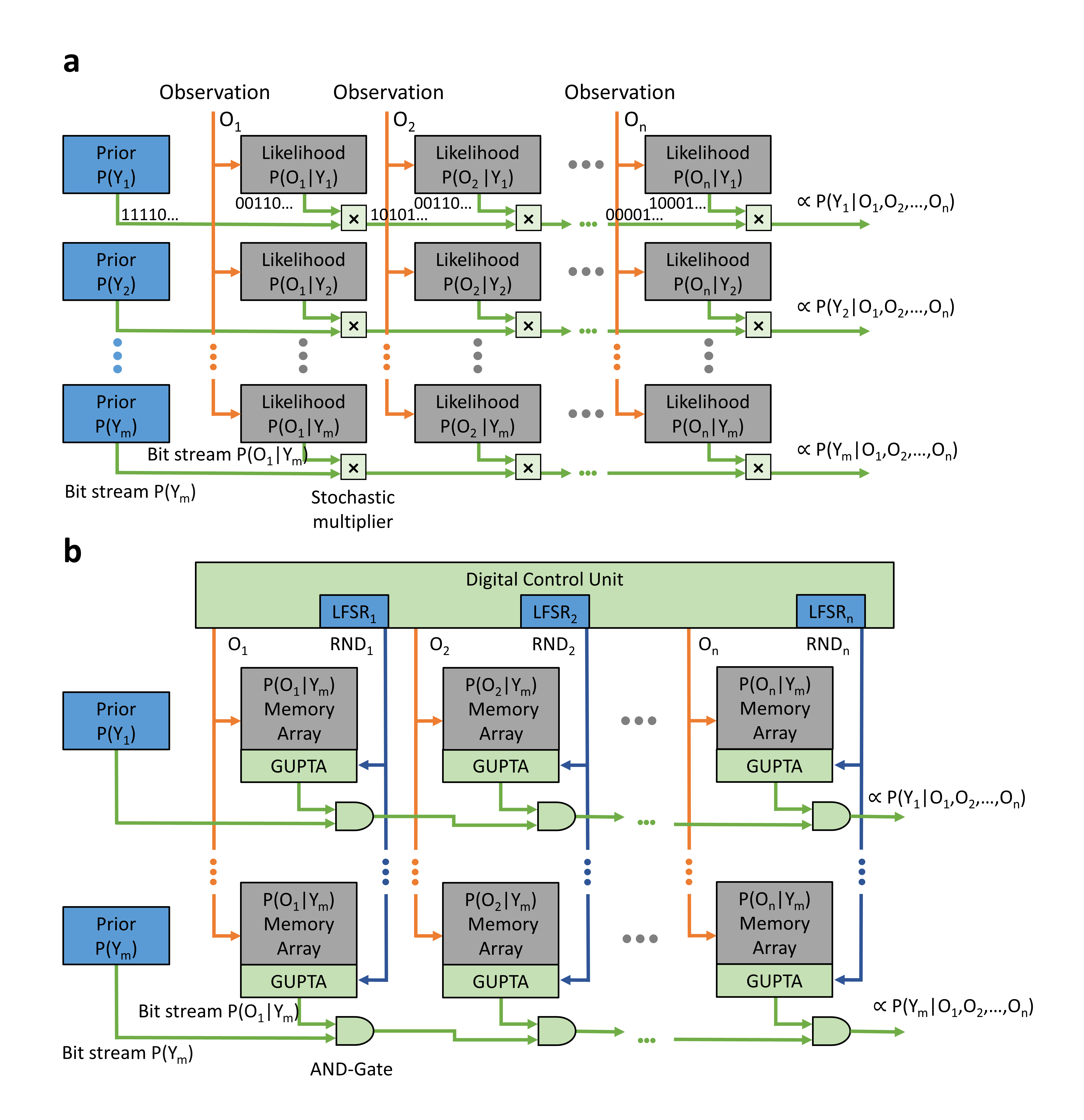}
\caption{
\textbf{General architecture of the Bayesian machine.}
\textbf{a} General architecture. The likelihoods are stored in likelihoods memory arrays (LMA) implemented by memristor arrays. Observations from the real world choose the appropriate probability value from LMAs, based on which proportional stochastic bit-streams are generated, which are multiplied by a stochastic multiplier. At the output, the bit-streams naturally encode the posterior distribution.
\textbf{b} Optimization of the Bayesian machine for hardware. Random numbers (RND) are generated using linear feedback shift registers (LFSRs), shared by column, and converted using digital ``Gupta'' circuits to a series of random bits proportional to the appropriate probability. Additionally, the likelihoods are normalized by the maximum likelihood value of the column to maximize the convergence speed of the machine. The stochastic multiplication is implemented by a single-bit AND gate.
}
\label{fig:cartoon}
\end{figure}

\begin{figure}[h!]
\centering
\includegraphics[width=\linewidth]{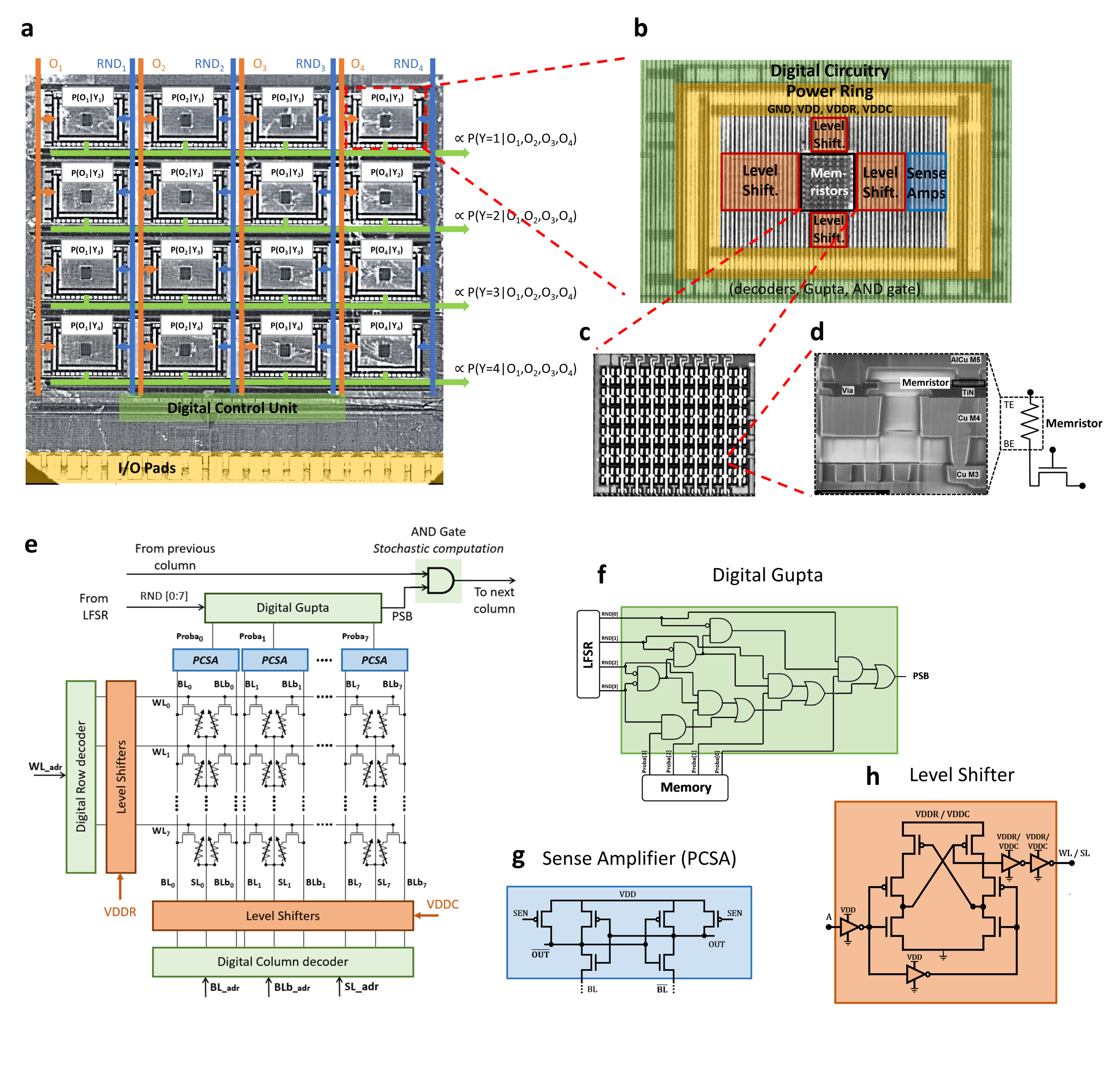}
\caption{\textbf{Fabricated memristor-based Bayesian machine.} 
\textbf{a}~Optical microscopy photograph of the Bayesian system die.
\textbf{b}~Detail of the likelihood block, which consists of digital circuitry and memory block with its periphery circuit.
\textbf{c}~Photograph of the 2T2R memristor array.
\textbf{d}~Scanning electron microscopy image of a memristor in the back end of line of our hybrid memristor/CMOS process.
\textbf{e}~Schematic of the likelihood block presented in \textbf{b}.
\textbf{f}~Schematic of the digital ``Gupta'' circuit \cite{gupta1988binary}, used for generating the proportional bit-stream (PSB) by comparing the read probability from memory with the random number generated by the LFSR. For compactness, the circuit is  presented here in a four-bits version. An eight-bits version is implemented on chip.
\textbf{g}~Schematic of the differential pre-charge sense amplifier used to read the binary memristor states.
\textbf{h}~Schematic of the level shifter, used for shifting nominal voltage input to forming and programming voltages of the memristors.
}
\label{fig:die}
\end{figure}

Bayesian inference aims at inferring a variable $Y$, based on a collection of observations $O_1$, $O_2$,...,$O_n$. For example, it might be used to detect the occurrence of a medical emergency, based on the status of a collection of sensors.
Bayesian inference produces a posterior distribution of the probabilities for the different possible values $y$ of $Y$, using
Bayes law: 
%
\begin{equation}
p(Y=y | O_1, O_2,...,O_n) \propto  p(O_1, O_2,...,O_n | Y=y) \times p(Y=y).
\label{eq:bayeslaw}
\end{equation}
%
$p(Y=y)$ is a prior distribution: the expected distribution of $Y$ in the absence of any observation.
For example, $Y=0$ might represent the absence of a stroke, $Y=1$ the occurrence of a minor stroke, and $Y=2$ the occurrence of a major stroke. Bayesian inference allows  evaluating the probability that a minor or major stroke is occurring, based on the probability  that such events happen at any time ($p(Y=y)$), and the likelihood factors $p(O_1, O_2,...,O_n | Y=y)$ that model the behavior of the sensors in the absence or presence of a stroke.

These likelihood factors $p(O_1, O_2,...,O_n | Y)$  feature a prohibitive memory cost, growing exponentially with the number of observations $n$. 
In real-life settings, they can be considerably simplified  
by considering the conditional independence of some observations. For example, suppose some sensors measure distinct aspects of the patients (e.g., heart rate and body temperature). In that case,  they may often be considered conditionally independent (meaning that once given the knowledge that a stroke is currently happening, the heart rate and body temperatures values can be regarded as  statistically independent processes).
For example, if all observations  are conditionally independent, Bayes law becomes
\begin{equation}
p(Y=y | O_1, O_2,...,O_n) \propto  p(O_1| Y=y) \times p(O_2| Y=y) \times  p(O_3| Y=y) ...  \times  p(O_n| Y=y)  \times p(Y=y),
\label{eq:naivebayeslaw}
\end{equation}
with a memory cost for the likelihood now growing linearly with $n$, and likelihood factors now becoming easy to model based on measurements of the sensor values, e.g., during the occurrence or in the absence of a stroke.

The memristor-based Bayesian machine implements equations such as eq.~\ref{eq:naivebayeslaw} in a topological manner. Each likelihood factor is implemented using independent memory arrays, and  multiplications are performed physically close to the memory arrays. The  multiplication result is then passed to the next memory array (Fig.~\ref{fig:cartoon}a). 
The observations $O_1$, ..., $O_n$ effectively act as addresses for the memory array, telling which likelihood value should be read. 

An important challenge of the memristor-based Bayesian machine is that multiplications are normally an area-expensive operation in CMOS, raising a concern if a multiplier is associated with each likelihood memory array.
For this reason, we rely on stochastic computing \cite{gaines1969stochastic,alaghi2013survey}.  
This computing paradigm encodes probabilities as  streams of random bits, where, at each clock cycle, the probability for the bit to be one is just the encoded probability.
The multiplication of probabilities can then be achieved  using simple AND gates,  with an extremely minimal area cost \cite{gaines1969stochastic}. An important aspect of this model of computation is that,  as in most practical settings, probabilities tend to be low, the output of stochastic computing AND gates is most cycle a zero value. Therefore, the different blocks of the memristor-based Bayesian machine  only need to pass single bits (Fig.~\ref{fig:cartoon}a) that are zeros most clock cycles: the memristor-based Bayesian machine limits data movement considerably.
Overall, due to this simplicity, the Bayesian machine just looks like a memory chip -- we call it a ``natively intelligent'' memory.

The architecture is here presented in the case where all observations can be considered conditionally independent, which is not always the case. In particular, when two redundant sensors measure the same phenomenon, they may never be regarded as independent in a good model, even conditionally to the inferred variable.  The Bayesian machine can also support this type of situation.
For example, if observations $O_2$ and $O_3$ are not conditionally independent,  it is  possible to pool them into a single column of the Bayesian machine, with memristor arrays storing joint likelihoods  $p(O_2,O_3| Y=y)$ (see Supplementary Note~1).

It should also be noted that in the case of a uniform prior, the prior blocks of the Bayesian machine may be removed entirely.

Beyond its conceptual elegance, the  design of the memristor-based Bayesian machine had to face several challenges.
First, stochastic computing requires random number generators. In our design, we use  linear-feedback shift registers (LFSRs) to generate uniformly randomly distributed numbers.  A single LFSR is used per column (Fig.~\ref{fig:cartoon}b): each row performing an independent stochastic computation, the different rows can rely on the same pseudorandom numbers. Then, at each clock cycle, each likelihood block needs to generate a random bit with the probability $p$ read from its likelihood memory array. For this purpose, each block features a simple digital circuit that takes as input the probability $p$ and the pseudorandom number generated by the LFSR to produce a pseudorandom bit with probability of $p$.
The schematic from this circuit is taken from \cite{gupta1988binary}, and we call it here a ``Gupta'' circuit.

Additionally, likelihoods in Bayesian models tend to be very low, and stochastic computing converges slowly when it is multiplying low probabilities \cite{winstead2019tutorial}. To optimize the operation of the Bayesian machine without any impact on its accuracy, we normalize the probabilities in a column so that the maximum likelihood in a column is one (see  Methods).

Another considerable challenge is that memristors are prone to errors. 
Industrial applications of memristors use strong formal error-correcting codes (ECC) \cite{chang2020envm}. 
Using ECC in the Bayesian machine is inappropriate, as error detecting and correcting circuits would dominate both area and energy consumption if they needed  to be replicated for each likelihood memory array \cite{gregori2003chip}. 
Therefore, we use an alternative strategy: memristors are used as single-level cells, and bits are programmed in a complementary fashion, and read differentially by precharge sense amplifiers comparing the resistance of two memristors (see Methods). This technique has been shown previously to reduce errors as efficiently as single error-correcting, double error-detecting codes (extended Hamming), using the same degree of memristor redundancy, and necessitating no error decoding circuit \cite{hirtzlin2020digital}. It is used here within a full system for the first time. The probabilities themselves are coded in binary representation as eight-bit integers (see Methods).

\subsection*{Characterization of a fabricated Bayesian machine}

\begin{figure}[h]
\centering
\includegraphics[width=1\linewidth]{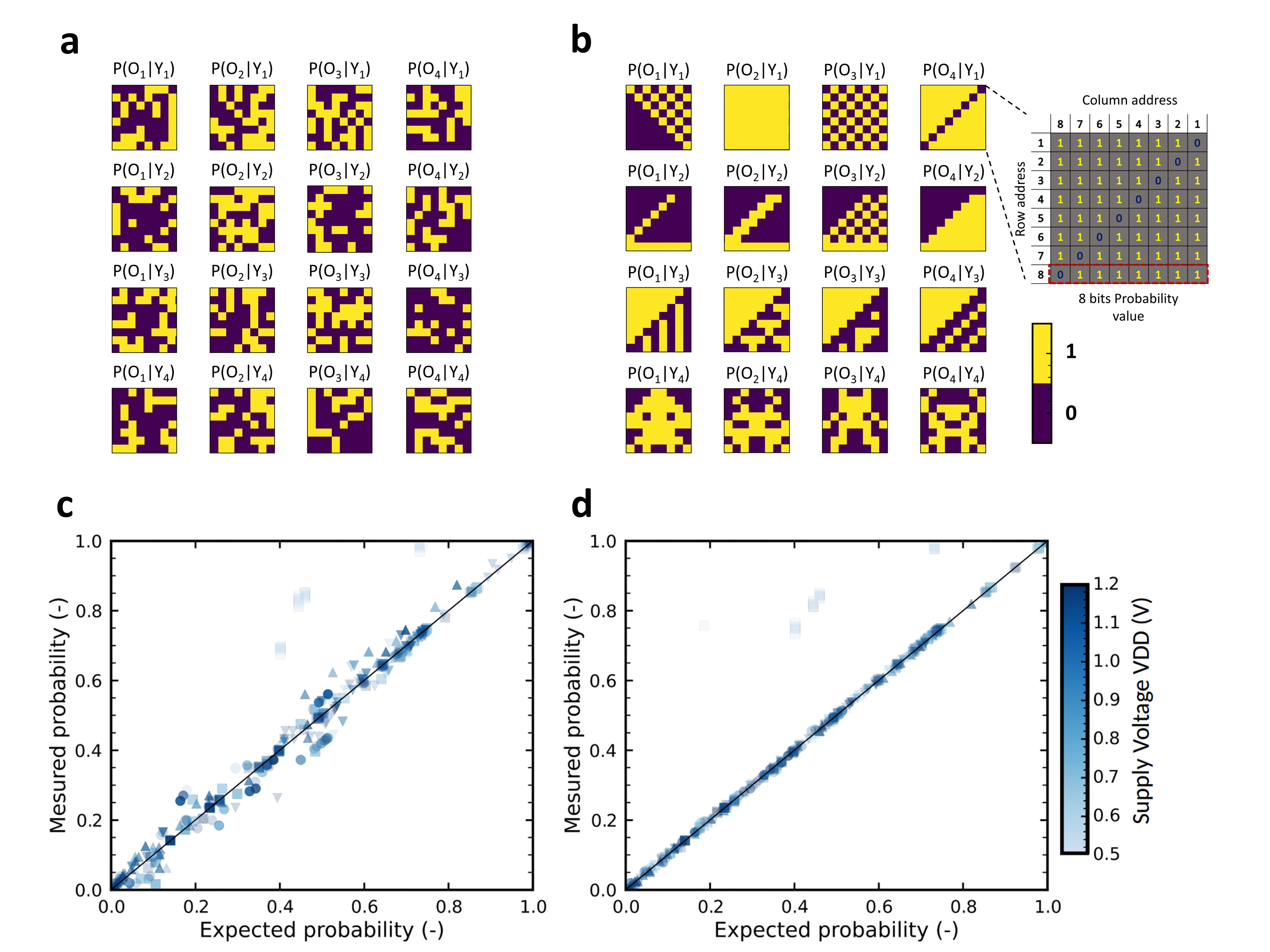}
\caption{
\textbf{Measurements of the fabricated memristor-based Bayesian machine.}
\textbf{a} Measurements of the likelihood stored in the memristors, before they have been formed. As the bits are programmed in a complementary fashion involving two memristors, the result of the measurement appears random.
\textbf{b} Measurements of the likelihood stored in the memristors, after they have been formed and programmed. No bit error is seen (see Suppl.~Note~3).
\textbf{c} Output of the Bayesian machine: measured posterior probability as a function of the expected value from Bayes' law. The different points correspond to random observation inputs. The different rows are pooled in the same graph. The points are obtained with various supply voltages VDD ranging between 0.5 and 1.2~volts. This graph is obtained with non-optimal LFSR seeds. The measured probabilities are obtained by averaging the experimental measurements over the full LFSR period (255 cycles).
\textbf{d} Same as \textbf{c}, using optimal LFSR seeds.
}
\label{fig:measurements}
\end{figure}

To validate the feasibility of memristor-based Bayesian inference,
we fabricated a prototype circuit in a hybrid CMOS/RRAM process. 
The CMOS part of the circuit is fabricated using a low-power foundry 130-nanometer process with four layers of metals (see Methods). The 130-nanometer node is a cost-effective sweet spot in terms of supported voltage levels, thanks to various MOS transistor options, which is appropriate for analog and mixed-signal low-power circuits with embedded memories,  targeting internet-of-things and edge applications. Hafnium oxide memristors are fabricated on top of the CMOS foundry layers, effectively taking the place of vias between metal layers four and five (see Fig.~\ref{fig:die}d and Methods).
This hybrid  process  features some constraints due to its partially academic nature: only four levels of metals are available for interconnection, the fifth level being used only to access memristors. Additionally, the design is probe tested. 
This test chip still allows us to demonstrate all the challenges associated with the fabrication of the Bayesian machine.
First, we need an original design flow enabling the distribution of the memristor blocks. Additionally, the system needs to distribute the higher-than-nominal voltages (up to five volts) required for forming and programming memristors. 
The nominal voltage of our foundry CMOS process is only 1.2~volts for the digital functions, whereas the forming operation of the  memristors, which is the most challenging in terms of voltage amplitude but occurs only once in the circuit lifetime, requires several volts. The distribution  of the higher-than-nominal voltages is a challenge in systems such as this one with massively distributed memory blocks that each need access to all voltage supplies.
Our test chip also allows verifying the error resilience of the design and its capability to perform inference at low voltage.
The system was designed with wide transistors and overall large safety margins, to allow the study of various programming regimes for the memristors.

Fig.~\ref{fig:die}a  shows the fabricated die with the superimposed structure of the Bayesian machine. The test chip implements a system with 16 likelihood memory arrays, organized in four rows and four columns (see Methods).  Fig.~\ref{fig:die}b shows the details of the likelihood memory arrays and their periphery circuitry (Fig.~\ref{fig:die}e shows the associated schematic). Fig.~\ref{fig:die}d shows an electron microscopy image of a memristor integrated into the back end of line of the die. Fig.~\ref{fig:die}g shows the differential precharge amplifier reading the memory bits,  Fig.~\ref{fig:die}h the level shifter capable of applying high voltages on the memristors based on nominal supply voltage inputs. The level shifters are used for forming and programming the memristors (see Methods).  Fig.~\ref{fig:die}f shows the digital Gupta circuits used for random bit generation. All subfigures of Fig.~\ref{fig:die} use consistent color codes. An important aspect of the design is that the whole core Bayesian machine does not necessitate access to the clock: the clock is only necessary in the digital control circuitry block outside the core machine, therefore limiting the energy cost of clock distribution.

Some results of the electrical characterization of the test chip are presented in Fig.~\ref{fig:measurements}.   Fig.~\ref{fig:measurements}a shows the measured results of reading the likelihood memory arrays of the chip, before forming and programming, naturally showing random values (as bits are stored in a complementary manner using two memristors, reading the memory with unformed bitcells leads to random results).  Fig.~\ref{fig:measurements}b presents the same measurements after programming: the intended  patterns are obtained without errors, showing the efficiency of the complementary programming technique (see Suppl.~Note~3). These artificial patterns were chosen so that performing Bayesian with random inputs allows exploring the whole range of possible output probabilities.

\FloatBarrier

The actual  operation of the Bayesian machine is shown in  Fig.~\ref{fig:measurements}c-d. In Fig.~\ref{fig:measurements}c, the LFSRs were initialized using random seeds. The x-axis represents the theoretical result expected from Bayes' law (i.e., the desired output for the Bayesian machine), while the y axis represents results measured experimentally by counting bits at the output at the die. The different points in the Figure are obtained by randomly changing the inputs $O_1$, $O_2$, $O_3$, and $O_4$ of the circuit, and by changing the nominal supply voltage $V_{DD}$. For each random set of input, the system was operated 255 clock cycles (i.e., the periodicity of the LFSRs). The number of ones outputted at each row is counted and divided by 255 to be converted as a probability, and plotted in the Figure. 

In Fig.~\ref{fig:measurements}c,  we see that the measured probabilities closely follow Bayes’ law, with some deviation. This deviation can be attributed to the imperfect nature of LFSR-generated pseudorandom numbers. The numbers generated on the different columns have correlations, which prevent stochastic computing from being perfectly accurate. 

Fortunately, this imperfection can  be avoided by an intelligent choice of the LFSR seeds. In the measurements of Fig.~\ref{fig:measurements}d, the chip was initially programmed by an optimal seed choice. We see that the measurements follow Bayes' law perfectly for all possible inputs, which highlights the high potential of stochastic computing for Bayesian inference.  Supplementary Note~2 details how the optimal LFSR seeds were chosen, their value, and the reason for their existence. This result shows that the  Bayesian machine is able to produce accurate outputs, despite its reliance on very simple pseudorandom numbers.

In addition to its limited data movement, which we analyze further in the next section, our approach offers two significant opportunities for low-energy operation. First, as the likelihoods are stored in non-volatile memristors (we observed no error in the stored likelihoods five months after they had been programmed, see Suppl. Note~3), the system provides an instant on/instant off feature: as soon as the power supply is turned on, the system is ready to perform Bayesian inference, without the need to load any data from memory. This feature means that unlike systems based on static RAMs, the power supply can be turned off any time the system is not used, without any penalty, offering an opportunity for energy saving.

Second, due to its fully digital nature, the system is flexible in terms of supply voltage.  Figs.~\ref{fig:measurements}c-d highlight that the system remains fully functional when reducing the power supply down to a value of 0.6~volts, although the nominal supply of our CMOS technology is 1.2~volts. This operation allows reducing power consumption by a factor of approximately four. At lower voltages (light blue points), the Bayesian inference becomes less accurate.
This voltage limit is due to the threshold voltage value of the thick oxide transistors used within the memory array,  around 0.6~volts. Even lower supply voltages could therefore be used by using lower threshold-voltage transistors.

\FloatBarrier

\subsection*{Energy efficiency of the memristor-based Bayesian machine}

\begin{figure}[h]
\centering
\includegraphics[width=0.75\linewidth]{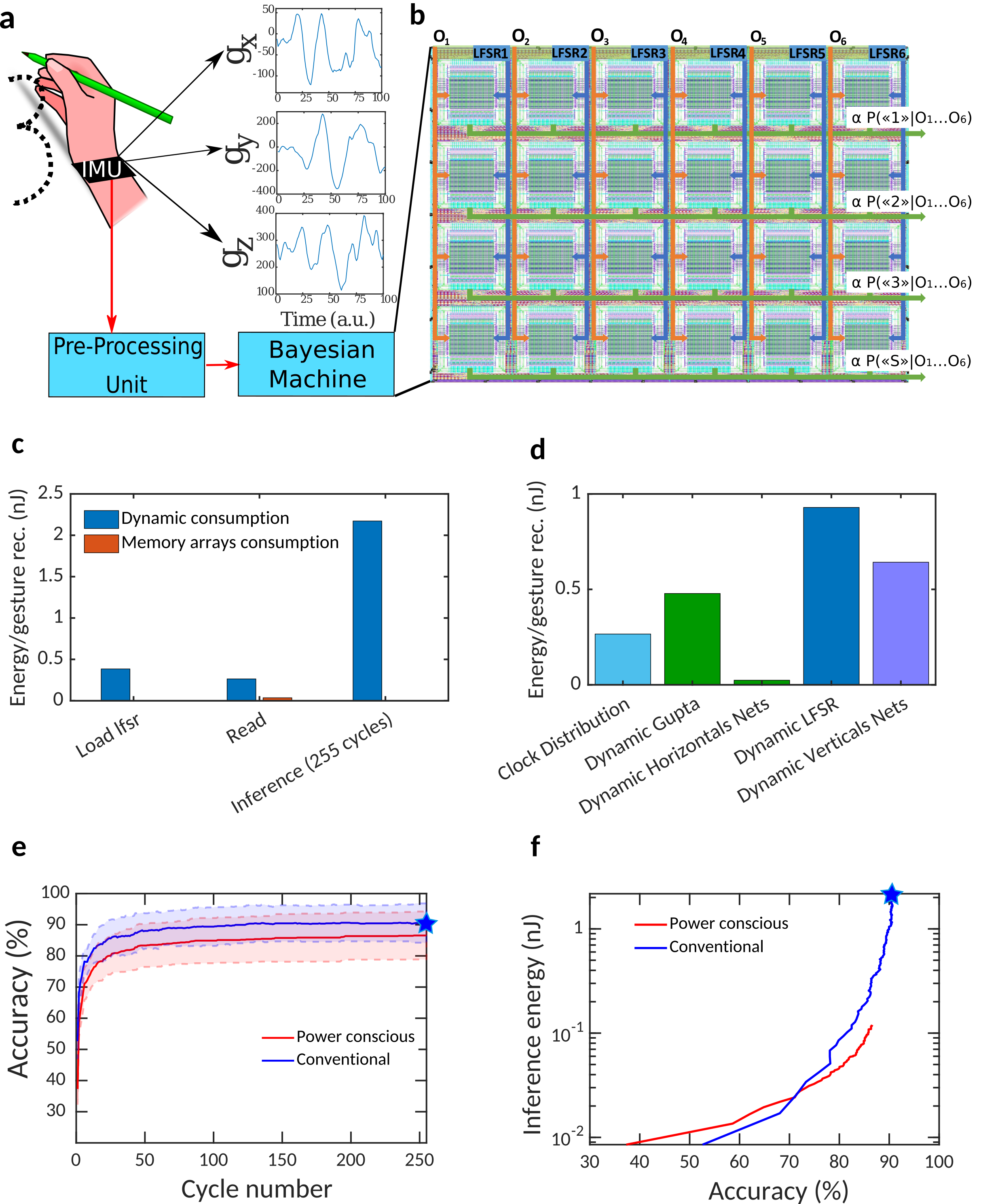}
\caption{
\textbf{Application of the Bayesian machine on a practical gesture recognition task.}
\textbf{a} Setup with inertial measurement unit used to record the gesture recognition dataset.
\textbf{b} Masks of the placed-and-routed Bayesian machine design used to perform the design-level gesture recognition analysis.
\textbf{c} Energy consumption of the system (Dynamic consumption and memory arrays) during the three phases of computing: loading the seeds into the LFSR, reading the memories, and the actual inference of 255 cycles. 
\textbf{d} Energy consumption of the system's important points during the inference phase for 255 cycles.
\textbf{e} Accuracy according to the number of cycle in the inference for two types of computation: using a ``power conscious'' method by taking into account only the first one out for the decision (in red) and using the conventional stochastic computing by using the maximum number of one out for the decision (in blue), and their respective STD intervals.
\textbf{f} Energy consumption  during the inference phase as a function of the accuracy for gesture recognition for the two methods. The stars correspond to the same point in both graphs \textbf{e} and \textbf{f}.
}
\label{fig:gesture}
\end{figure}

Our test chip allows validating the possibility to address the challenges of designing and fabricating the memristor-based Bayesian machine. This system is, however, not adapted to evaluate the power consumption of a final system, as the test chip is too small to implement real-life applications. Additionally, the constraints of the semi-academic process (notably probe testing involving high capacitive loads on the outputs, see Methods) and the wide transistors that we employed cause a high increase of the dynamic capacitive energy consumption. 
To evaluate power consumption, we switch to a larger design, and a realistic application, and use industry-standard integrated circuit design tools to evaluate energy consumption with a fine granularity.

We focus on an application of gesture recognition. The input to the Bayesian machine is a selection of features extracted from the time traces on an inertial measurement unit (IMU, see Methods). The goal of the system is to recognize the gesture performed by a user wearing the IMU (see Fig.~\ref{fig:gesture}a): 
the gesture of writing the digit one, the digit two, the digit three, or a signature (see Methods). 
This task is performed by a scaled-up version of the Bayesian machine, using 24 (six columns, four rows) four-kilobits likelihood memory arrays.

We designed and laid out this system in our reference process and evaluated its energy usage. 
We can see in the  image of the masks shown in Fig.~\ref{fig:gesture}b) that in this scaled design, the area of the memristor arrays is now dominant, with regards to the memory periphery circuitry and the wiring of the Bayesian machine.
The energy consumption is based on an exact scenario using value change dump files, and required adapting the standard flow of energy analysis, which is not naturally adapted for systems where the memory is as distributed as ours (see Methods). The energy consumption of the memory arrays and  the digital circuits is evaluated independently using circuit (Spice) simulation and digital circuits analysis tools. In both cases, parasitic capacitances were extracted based on a complete layout and were included in the energy analysis (see Methods).  

Fig.~\ref{fig:gesture}c shows the energy consumption of the different elements of the system in the three operation phases (after the likelihoods have been programmed). The LFSR initialization consists in loading the seeds of the six LFSRs of the circuit. This operation consumes 0.38~nanojoules; it needs to be performed once when the system is turned on, and does not need to be repeated as long as the power remains on. It, therefore, remains a minor contribution to energy consumption. This energy could also substantially be reduced by hardwiring the value of the optimal seeds  (whereas seeds are loaded from external inputs in our design, see Methods). 
By contrast, the memory read operation needs to be performed each time a new input is presented to the system, and consumes a total of 0.3~nanojoules, including both the energy associated with the memory circuit themselves and the digital control circuitry. The actual stochastic inference, corresponding to the stochastic computation, consumes 2.2~nanojoules (assuming that all 255 cycles of the LFSRs have been operated). Therefore, in sharp contrast with von Neumann-type architectures \cite{pedram2017dark}, the energy consumption of the computation is dominant with regards to the energy for accessing data, highlighting the benefits of computing close to memory.

To further analyze the energy usage of our architecture, Fig.~\ref{fig:gesture}d details the different sources of energy consumption during the stochastic computation. The clock distribution represents 11\% of the energy consumption. This number remains relatively modest, because the clock is not distributed within the Bayesian machine itself, but only to the external digital control circuitry. 
The computation itself (AND gates) and the transfer of the output of the blocks to the next block, through horizontal wires, represent only 1\% of the total energy consumption. Random number generation  (LFSRs, Gupta circuits)  represents 60\% of the energy consumption, and their distribution  through vertical wires 28\%.
Future efforts should therefore focus on this part (see Discussion).

Before that, an obvious technique for reducing energy consumption is to reduce the number of cycles computed during stochastic inference. This reduction naturally impacts accuracy, as highlighted in Fig.~\ref{fig:gesture}e. This Figure shows, for the gesture recognition task, the accuracy of the Bayesian machine, as a function of the number of considered cycles. Numbers higher than 255 serve no purpose, as 255 is the periodicity of the eight-bits LFSRs. 
We consider the traditional stochastic computing strategy, as well as a ``power-conscious'' strategy. 
In the traditional approach, the system is operated for a fixed number of cycles, and the recognized gesture is chosen as the output that generated the highest number of ones.
In the simplified power-conscious strategy,  computation is stopped as soon as any of the circuit outputs  produces a one, and this output gives the  recognized gesture. 
We see that in both cases, cycles numbers as low as 50 allow approaching the accuracy obtained with 255 cycles.
Based on these results,
Fig.~\ref{fig:gesture}f shows the interplay between accuracy and energy consumption, using both strategies. The power-conscious accuracy is most favorable only for accuracy higher than 70\% and smaller than 90\% (which is the maximum accuracy that can be reached on this particular dataset).  For higher and lower accuracies, conventional stochastic computing is more favorable. Overall reducing the number of cycles appears a highly effective strategy: accepting an accuracy reduction of only one percentage point allows reducing the energy consumption by a factor 2.9.

To benchmark the energy efficiency of our approach, we also implemented the Bayesian gesture recognition task on a microcontroller unit (MCU), with an optimized approach using integer computation solely  (see Methods). MCUs are tiny computers incorporating all their logic, volatile, and non-volatile memory on a single chip. They are currently the mainstream approach for providing AI at the edge in energy-constrained contexts \cite{warden2019tinyml}. Experimental measurements (see Methods) showed that recognizing one gesture with the MCU used ten microjoules, i.e., 5,000 times more energy than the memristor-based system, even in the more energy-consuming strategy using the 255 LFSR cycles is chosen.

\FloatBarrier

\section*{Discussion}

Our results show that a Bayesian machine can be implemented in a system with distributed memristors, performing computation locally, and with minimal energy movement, allowing the computation of Bayesian inference with an energy efficiency more than three orders of magnitude higher than a standard microcontroller unit. Due to its reliance on non-volatile memory, and its sole use of read operation,  once the likelihoods have been programmed, the system may be powered down anytime while regaining functionality instantly. It may also be operated at low and possibly varying supply voltage.
While Bayesian models are usually considered computationally expensive, these results suggest that complex models might be embeddable  at the edge, with low power consumption. 
This prospect could allow edge systems to benefit from the qualities of Bayesian inference to deal with highly uncertain situations with little data, and to make predictions using an explainable mode. This feature could be particularly useful for medical devices, or for monitoring  difficult environments. This second application is particularly interesting,   as an additional benefit of stochastic computing is that our system is naturally resilient to soft errors: bit errors can make one cycle wrong, but will be averaged throughout the computation. 
As memristor storage is also more resilient to radiation than static RAM \cite{petzold2019heavy}, this feature can make the Bayesian machine appropriate for extreme environments.

All our results were obtained using a 130-nanometer process, showing that outstanding energy efficiency may be obtained even using an inexpensive technology. As the energy consumption is dominated by digital circuitry, it could be considerably reduced by scaling the design to more aggressive technology nodes.

Our energy analysis revealed that 88\% of the energy consumption during the inference phase was due to random number generation and distribution. The  generation cost is due to the use of LFSRs, and the  distribution cost is due to the non-local nature of random number generation (our system used a single LFSR per column, shared by all the likelihood blocks of the column). In our vision,  this energy should be reduced by again relying on nanodevices.
Stochastic nanodevices can generate high-quality random bits locally, at a very low area and energy cost \cite{vodenicarevic2017low,borders2019integer}.

In conclusion, this paper highlights that Bayesian inference might be an exciting way to bridge intelligence at the edge, due to the reliance on local nanotechnology-based memories, and opens the way for the body of research, going from the algorithmic developments of edge Bayesian inference, to device, circuits, and architectural developments.

\section*{Acknowledgements}
This work was supported by European Research Council starting grant NANOINFER (reference: 715872).
The authors would like to thank A.~Cherkaoui, M.~Faix, R.~Frisch, J.~Grollier, L.~Herrera-Diez, E.~Mazer, J.~Simatic, and S.~Tiwari   for discussion and invaluable feedback.

\section*{Author contributions statement}
K.E.H and T.H. designed the test chip, under the supervision of J.M.P and D.Q. J.M.P. designed the mixed-signal circuits of the test chip. M.B. and T.H.  performed the electrical characterization of the system. K.E.H. and C.T. designed the scaled up version of the system. R.L. developed the gesture recognition application, and C.T. adapted it to the memristor-based Bayesian machine. J.D. and P.B developed the initial theory of the Bayesian machine. E.V. led the fabrication of the test chip. D.Q. supervised the work and wrote the initial version of the manuscript.
All authors discussed the results and reviewed the manuscript. 

\section*{Competing interests}
The authors declare no competing  interests.

\section*{Data availability}
The datasets analysed and all data measured in this study are available from the corresponding author upon reasonable request.

\section*{Code availability}
The software programs used  for modeling the Bayesian machine are available from the corresponding author upon reasonable request.

\section*{Methods}

\subsection*{Fabrication of the system}

The CMOS part of our  test chip is fabricated using a low-power foundry 130-nanometer process with four layers of metals. The memristors are fabricated on top of exposed vias, and are composed of a TiN/HfO$_x$/Ti/TiN stack. The active HfO$_x$ is deposited by atomic layer deposition and is 10~nanometers thick. The Ti later is also 10~nanometers thick, and the memristor structure has a diameter of 300~nanometers. On top of the memristors, a fifth layer of metal is deposited. The input/output pads are aligned in a line of 25~pads designed for characterization by a custom probe card.

\subsection*{Design of the demonstrator}

The memristor-based Bayesian system is a hybrid CMOS/nanotechnology integrated circuit, with memory arrays embedded within the logic. 
In the absence of a foundry design kit supporting such designs, and because our design required the use of three different supply voltages,
we relied on a homemade semi-automated design flow. 

The supply voltage VDD, with a nominal value  of 1.2~volts, is used to power computation logic and sensing circuitry. The supply voltage VDDR, with a maximum allowed value of 5~volts,  supplies the memory rows drivers, which feed the selection transistors gates called word line (WL, see Fig.~\ref{fig:die}e). Finally, the supply voltage VDDC, also with a maximum possible value of 5~volts,  supplies the memory columns drivers, which feed the Bit Lines (BL) or the Source Lines of the memory array (see Fig.~\ref{fig:die}e). The whole system uses  a single common ground (GND) for all voltage domains. The memristor memory array is a 64 bit cell array. The bit cell is implemented by a two-memristor, two-selection-transistor (2T2R) structure. The n-type selection transistors, as well as the level shifters (see Fig.~\ref{fig:die}h), driving the memristor arrays are based on thick-oxide transistors supporting voltages  up to five volts. This high-voltage feature is necessary for forming and programming the devices. Each column of the array features a precharge sense amplifier (PCSA) as  sensing circuitry (see Fig.~\ref{fig:die}g), designed using low-voltage high-threshold transistors with thin gate oxide, chosen to minimize energy consumption during sensing operation.
The memristors arrays and their mixed-signal peripheral circuitry were designed, placed and routed by hand using the Cadence Virtuoso electronic design automation (EDA) tool. 
The Siemens Eldo simulator was used for analog simulations. 

By contrast, all digital computation blocks (decoders, LFSR, Gupta, registers...) were described using the SystemVerilog hardware description language, logically verified using the Cadence NC-Verilog Simulator, synthesized using the Cadence Encounter RTL Compiler, then placed and routed independently using the Cadence Encounter RTL to GDSII tool, following a semi-automated flow developed by the foundry.  All digital circuits use the thin gate oxide high-threshold transistors.

The layouts of computation blocks, along with the layouts of memory blocks, were placed and routed manually in a full-custom fashion. All physical verifications (design rule check, layout-versus-schematic comparison, and antenna effects design rule checks) of the final design were performed using dedicated Calibre EDA tools.

\subsection*{Measurements of the system}

Our system is probe tested using a custom-made 25-pads probe card, connected to a custom dedicated printed circuit board (PCB) featuring level shifters and other discrete elements by SubMiniature A (SMA) connectors. The PCB  connects inputs and outputs of our test chip to an ST Microelectronics STM32F746ZGT6 microcontroller unit, two Keysight B1530A waveform generator/fast measurement units, and a Tektronix DPO~3014  oscilloscope. The microcontroller is connected to a computer using a serial connection, while other equipments are connected to the computer using a National Instruments GPIB connection. The tests are then operated using python within a single Jupyter notebook controlling the whole setup.

Before the operation of the Bayesian machine, the memristors need to experience a unique ``forming'' operation, to create conductive filaments.
Forming is realized memristor-by-memristor. VDDC is set to 3.0 volts,  VDDR to  3.0 volts, and VDD to 1.2 Volts. Using the digital circuitry of our test chip, a programming pulse is applied to each memristor during one microsecond.

Once formed, memristors can be programmed in low-resistance or high-resistance states (LRS or HRS). To program in the LRS, a SET mode is activated. VDDC is set to 3.5~volts,  VDDR to  3.0~volts,  and VDD to 1.2~volts. Using the digital circuitry of our test chip, a programming pulse is applied to each memristor to be programmed in LRS during one microsecond. To program in the HRS, a RESET mode is activated. VDDC is set to 4.5~volts,  VDDR to  4.9~volts,  and VDD to 1.2~volts. Using the digital circuitry of our test chip, a programming pulse is applied to each memristor to be programmed in HRS during one microsecond. The polarity of this pulse is opposed with regard to the SET pulse.

Memristors are always programmed in a complementary fashion, following the principle established in \cite{hirtzlin2020digital}. 
The 2T2R bit cell contains two memristors. To store a zero, the left memristor (along BL) is programmed to HRS and the right memristor (along BL$_b$) to HRS. Conversely, to store a one, the left memristor is programmed to LRS and the right memristor to HRS.
No program-and-verify strategy is employed. 
The likelihoods are stored as eight-bits integers proportional to the likelihood values (see Fig.~\ref{fig:measurements}b). They are normalized by the maximum value of the likelihoods of a column (i.e., the maximum  likelihood value is stored as integer FF, in hexadecimal representation). This normalization allows stochastic computing to converge faster, as explained in the main text.

Once the likelihoods have been programmed, high voltages are no longer needed, and the test chip can be used in its normal mode to perform Bayesian inference based on observations. All three supply voltages are set to the inference supply voltage, normally 1.2~volts, but which can be reduced up to 0.5~volts. The outputs of the Bayesian machine are recovered by the microcontroller unit and transmitted to the computer.

\subsection*{Gesture recognition task}

The gesture recognition task is realized on a  dataset collected in-lab, including ten subjects. Each subject was asked to perform four gestures (writing the digits one, two, three, and a signature specific to each person) in the air. The subjects were not given any instruction on how to perform the gestures, leading to a high diversity within the dataset. The gestures were recorded using the three-axis accelerometer of a standard inertial measurement unit. Each subject repeated the same move between 25 to 27 times. The recording time varied by subject and gesture, and ranged from 1.3 to 3~seconds. 
We extracted ten features, named $F_0$ to $F_9$, from each recording, after filtering of the gravity (see Supplementary Note~4 for the associated equations): mean acceleration, maximum acceleration on the three axis, variance of the acceleration on the three axis, mean value of the jerk of the acceleration, and maximum value of the jerk of the acceleration on the three axis.

We train the system using 20 of the 25-27 recordings for each subjects, and the last 5-7 recordings are used to test it.  Our model assumes a uniform prior, and the resulting machine, therefore, features no prior block.
Training consists in adjusting the likelihoods $p(O_n| Y=y)$ to the training data.  As the training data is very limited, we fitted these likelihoods by Gaussian functions (using the fitdist MathWorks MATLAB  function). We then discretized the resulting Gaussian distributions to the 512 possible input values of the observations in the scaled-up Bayesian machine. We then normalize the probabilities in each column of the Bayesian machine so that the maximum value is one (see Main text). Finally, we  quantified the normalized probability values to eight-bit integers, with the value zero equivalent to 1/256 and 255 to 256/256.   

To optimize further the energy consumption of the system, we use  only six of the extracted ten features in the Bayesian machine.  Based on a systematic study, the features deleted for the experiment are $F_0$, $F_2$, $F_3$, $F_8$. 
Additionally, we realized that broadening the Gaussians obtained when fitting the data allowed stochastic computing to converge faster, allowing the system to reach better accuracy.  Therefore, in all our results, the standard deviation of the Gaussian in the fitted likelihoods is multiplied by a broadening coefficient of 1.3 with regards to the initial fit.

\subsection*{Simulation of scaled Bayesian machines}

The scaled-up Bayesian machine features six columns and four rows of likelihood blocks. Each likelihood block features an array of 128$\times$64 memristors arranged using the same differential structures as our test chip, therefore implementing four kilobits per array.
We developed a behavioral MathWorks MATLAB model of the machine,  a synthesizable SystemVerilog description, and test benches for both models using consistent input files. Both models were verified to be equivalent for all possible inputs and for all cycles. We synthesized the SystemVerilog description and placed and routed the whole Bayesian machine in our reference technology (see Fig.~\ref{fig:gesture}b). Post-place-and-route simulation, including the delays due to the gates and the parasitic capacitances, gave results that still matched perfectly the MATLAB model of the Bayesian machine.

\subsection*{Energy consumption estimates}

The energy estimates on the scaled design were obtained using a hybrid methodology. These estimates focus on the inference phase, i.e., the actual operation of the Bayesian machine when various inputs are presented, after the memristors have been formed and the likelihoods programmed.
The energy consumption of the memristor arrays themselves are obtained using circuit simulations (based on the Siemens Eldo simulator), including the parasitic capacitance extracted from the memory array layout. 
The energy consumption of the rest of the system is obtained using the Cadence Voltus power integrity solution framework. These estimates use value change dump (VCD) files obtained from our test bench, ensuring that the energy estimates correspond to a realistic situation.

A challenge of these estimates is that when estimating energy consumption, it is crucial that Cadence Voltus models the behavior of the memristor arrays properly, as the energy consumption of the system depends directly on the output of these blocks. However, as they are custom blocks, and not included in the standard library of the foundry, special developments were required. We programmed, using MATLAB, a memristor array to liberty file compiler, providing, based on the likelihoods programmed in a memory block, a file describing to Cadence Voltus the functionality of the array. During the MATLAB simulation, we extract the intermediates values of the inference, and we include them as a memory output. For that purpose, we create a new liberty file that will be used in the place-and-route operation. This liberty file specifies the output for each memory as a function of the input and addresses. This method can only be used to estimate the energy consumption during the inference phase as the outputs are tailor-made for this phase.

\subsection*{Implementation of the functionality of the Bayesian machine on a microcontroller unit}

As a benchmark, we also implemented the gesture recognition task on an ST Microelectronics STM32F746ZGT6 microcontroller unit (MCU, integrated on a test Nucleo-F746ZG board). This type of MCU is commonly used for edge artificial intelligence. 
Our implementation was programmed in the C language using the ST Microelectronics STM32 Cube integrated development environment and was optimized for running on the MCU (the stochastic computation of our test chip is replaced by standard integer addition, and the likelihoods were replaced by log-likelihoods to avoid multiplications). 
To perform the benchmark, the MCU computes gesture recognition for all possible inputs sequentially and blinks an LED on the board every one-million inference to allow precise timing. The energy consumption of the board was measured using a standard Ampere meter.

\bibliography{sample}

\end{document}